**Real-space imaging reveals symmetry-selected nonlinear energy routing in a mechanical resonator**

Ya Zhang[1,a)], Yuko Terasawa[1], Qian Liu[1], Shumpei Takenaka[1], Hua Li[1,2], Yutao Xu[3], Xueyong Wei[3], and Kazuhiko Hirakawa[1,4]

[1]*Institute of Engineering, Tokyo University of Agriculture and Technology, 2-24-16 Koganei-shi, Tokyo, 184-8588, Japan*

[2]*Shanghai Institute of Microsystem and Information Technology, Chinese Academy of Sciences, Shanghai 200050, China*

[3]*School of Instrument Science and Technology, Xi'an Jiaotong University, Xi'an 710049, China.*

[4]*Institute of Industrial Science, University of Tokyo, 4-6-1 Komaba, Meguro-ku, Tokyo 153-8505, Japan*

Nonlinear energy routing among modes underlies phenomena ranging from internal resonance and wave mixing to frequency-comb generation, yet modal interactions are typically inferred from spectra rather than imaged in real space. This leaves unresolved how energy is spatially routed and what determines which pathways are selected. Here, we use phase-locked multi-harmonic stroboscopic interferometry to reconstruct harmonic-resolved differential displacement maps in a nearly mirror-symmetric microelectromechanical resonator. These maps reveal that harmonics generated by a driven mode can be carried by distinct spatial eigenmodes, directly resolving pathways of nonlinear energy transfer. We further show that such mode-selective routing occurs even away from integer frequency matching: generated harmonics are dominated by eigenmodes sharing the driven mode's mirror parity, whereas spectrally closer opposite-parity modes remain strongly suppressed. A nonlinear modal framework links this hierarchy to symmetry-dependent modal-overlap integrals. These results identify spatial symmetry as a selection rule for nonlinear energy routing.

---

a) Electronic mail: zhangya@go.tuat.ac.jp

## Introduction

Nonlinear energy routing among modes underlies a wide range of phenomena in mechanical, optical, and quantum systems. In resonant structures, energy is not confined to a single eigenmode but can be redistributed among multiple modes through nonlinear interactions [1-4]. When modal frequencies approach integer ratios, internal resonance enables coherent energy exchange between distinct degrees of freedom [5-13]. Such nonlinear interactions profoundly modify resonator dynamics, enabling frequency stabilization [14, 15], synchronization [16], enhanced sensing [17, 18], control over energy-dissipation pathways [19, 20], vibrational energy harvesting [21], and frequency-comb generation [22-24]. While frequency-commensurate interactions have been widely investigated in micro- and nanoelectromechanical resonators, a more fundamental question remains unresolved: beyond frequency matching, what determines whether a model energy-transfer pathway is allowed or suppressed?

Experimentally, nonlinear modal interactions are typically inferred indirectly from spectral signatures, such as amplitude bifurcation, frequency pulling, and sideband generation. Although these measurements identify nonlinear response and the frequencies generated by nonlinear motion, they provide no direct information about how vibrational energy is spatially reconfigured within the resonator. The observation of a spectral harmonic alone cannot determine whether that component reflects a temporal distortion of the driven mode, a resonance-amplified response of another eigenmode, or off-resonant redistribution into a distinct spatial mode. Recent spatial mapping of strongly driven membranes has reported spatial distortions interpreted by localized overtones [25] or interferometric readout nonlinearity [26]. However, the model energy-transfer pathways have not been resolve. Addressing this problem therefore requires spatial imaging with controlled readout nonlinearity to determine whether nonlinear routing is governed by frequency proximity or by symmetry-based selection.

Symmetry plays a central role in determining admissible interactions across physics. In optical and quantum systems, transition amplitudes and nonlinear interaction strengths are often governed by symmetry-induced cancellation of matrix elements, giving rise to well-known selection rules [27-30]. By analogy, nonlinear modal coupling in mechanical resonators should also be constrained by spatial symmetry through the vanishing of modal-overlap integrals. Although symmetry dependence is implicit in modal-overlap formulations of nonlinear vibrational dynamics, whether such symmetry constraints select the spatial pathways of nonlinear energy transfer has not been established experimentally.

Here, we use phase-locked multi-harmonic stroboscopic interferometry to reconstruct harmonic-resolved differential displacement maps in a nearly mirror-symmetric mechanical resonator. These maps resolve the spatial eigenmode content of individual harmonic components, directly revealing the spatial pathways through which nonlinear energy is routed. This allows us to distinguish temporal distortion of the driven mode from nonlinear redistribution into distinct spatial eigenmodes.

We first establish this capability at a frequency-commensurate internal resonance. At a 1:5 internal resonance, the fifth harmonic generated by fundamental-mode excitation is carried by the third bending eigenmode, directly resolving the spatial pathway of resonantly enhanced modal energy transfer. We then show that such mode-selective routing occurs even away from integer frequency matching. In the off-resonant regime, generated harmonic components are dominated by eigenmodes sharing the driven mode's mirror parity, whereas spectrally closer opposite-parity modes remain strongly suppressed. A nonlinear modal framework based on geometric nonlinearity shows that the relevant cubic coupling coefficients are governed by symmetry-dependent modal-overlap integrals, identifying mirror parity as the selection rule governing nonlinear energy routing.

By directly imaging the spatial pathways of nonlinear modal energy transfer, this work separates two roles that are often conflated in nonlinear resonators. Frequency commensurability provides resonant amplification of existing coupling channels, whereas symmetry-determined modal overlap selects which channels are intrinsically favored or suppressed. Together, these results identify spatial symmetry as a selection rule for nonlinear energy routing and provide a method to symmetry-engineered control of energy flow in multimode nonlinear wave systems, including micro- and nanomechanical resonators, optomechanical platforms, and phononic structures.

## Results

### Modal landscape and parity classes

To establish the modal landscape required for probing symmetry-selected nonlinear energy routing, we employ a doubly clamped silicon-on-insulator resonator fabricated using microelectromechanical-systems (MEMS) technology. Figures 1A–C show the optical micrograph, device schematic and measurement setup. A weak structural asymmetry is intentionally introduced to facilitate excitation of antisymmetric modes while preserving approximate mirror symmetry about the beam midpoint. The beam is driven inertially by an external piezoelectric actuator ($V_{AC}$) and measured by either multi-harmonic stroboscopic interferometry [31, 32] or by laser Doppler vibrometry (LDV). The midpoint defines the approximate mirror plane (dashed line in Fig. 1A), enabling classification of vibrational modes according to spatial parity.

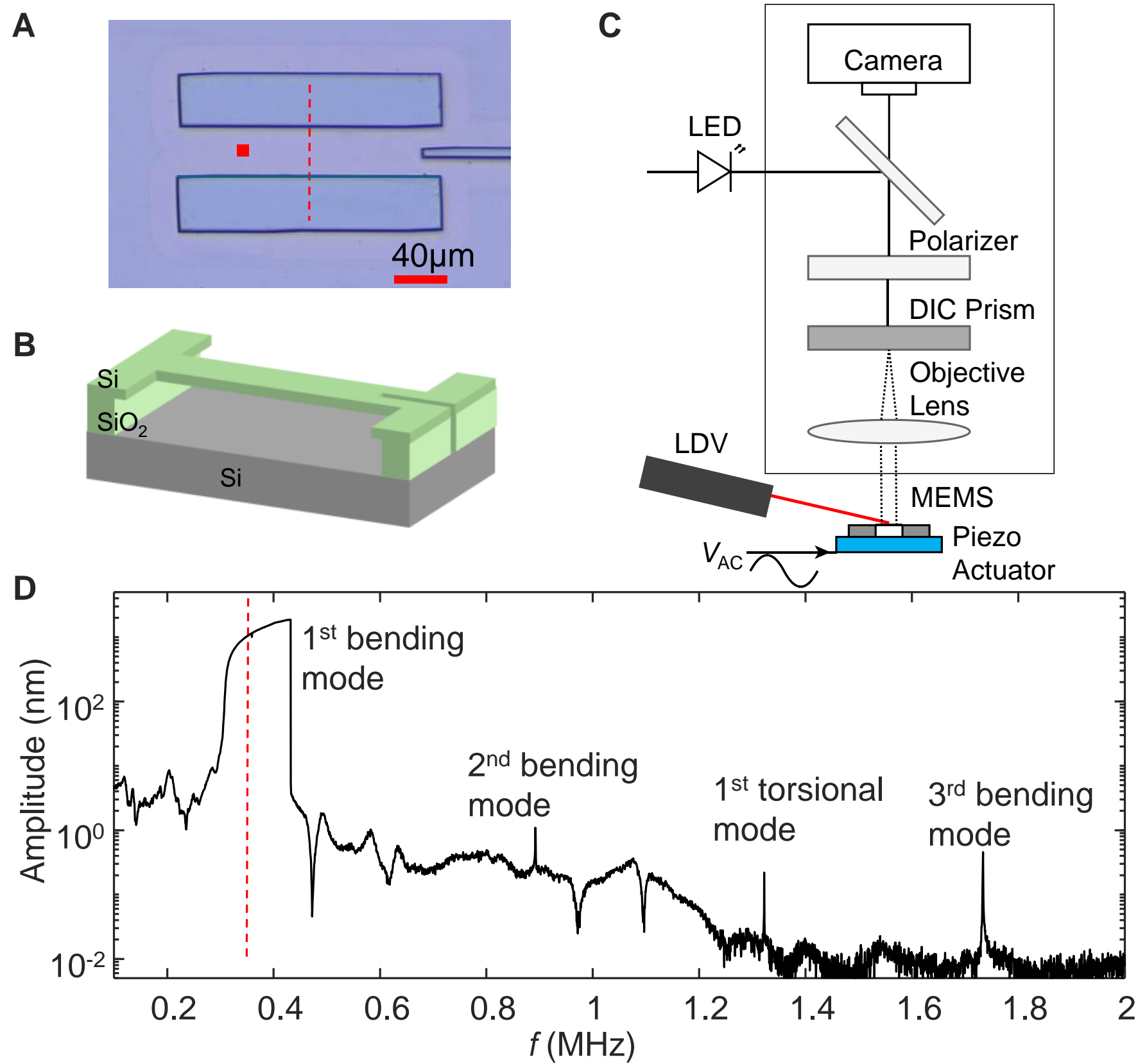

**Figure 1. Device architecture, measurement scheme, and broadband modal spectrum.** (**A**) Optical micrograph of the doubly clamped SOI mechanical resonator used in this work. (**B**) Schematic structure of the resonator. (**C**) Schematic of the measurement setup. The beam is driven inertially by an external piezoelectric actuator with an AC voltage $V_{AC}$, and its out-of-plane

displacement is measured either by phase-locked multi-harmonic stroboscopic interferometry or by laser Doppler vibrometry (LDV). (**D**) Broadband frequency response measured under forward frequency sweep from 0.1 to 2 MHz with $V_{AC} = 10$ V.

The broadband frequency response (0.1–2 MHz) measured under forward frequency sweep is shown in Fig. 1D. The actuation amplitude is set sufficiently high ($V_{AC} = 10$ V) to resolve multiple resonances within a single scan. Under these conditions, the first mode exhibits pronounced Duffing-type hardening, extending its resonance branch from approximately 311 kHz to 440 kHz and producing characteristic jump phenomena and bistability. Higher-order modes remain closer to the linear regime because inertial actuation efficiency decreases rapidly with increasing mode order. Distinct peaks corresponding to the first three bending modes and one torsional mode are identified in Fig. 1D. Due to the combined effect of inertial driving and approximate mirror symmetry, mirror-symmetric bending modes are excited more efficiently than mirror-antisymmetric modes, as reflected in their relative response amplitudes.

Finite-element simulations of the first six linear bending eigenmodes are shown in Fig. 2. For each mode, both the displacement profile $\phi$ and the corresponding spatial derivative $\phi'$ are presented. This distinction is important because the stroboscopic DIC technique measures spatial differential displacement, which is proportional to the local slope of the mode shape. With respect to the beam midpoint, the first, third and fifth bending modes (odd modes) are mirror-symmetric for the displacement profile, whereas the second, fourth and sixth bending modes (even modes) are mirror-antisymmetric. This symmetry classification defines two distinct modal families that form the basis for testing symmetry selection in nonlinear modal interactions in the following sections.

The experimentally observed resonance peaks in Fig. 1D are assigned by matching both resonance frequencies and spatial profiles. Importantly, the nonlinear frequency tuning of the fundamental mode naturally spans the approximate commensurability condition $f_1 \approx f_3/5$ (Fig. 1D). The corresponding 1:5 relation, $f_3/5$, is indicated by a red dashed line in the spectrum. This spectral overlap establishes the prerequisite for 1:5 internal resonance, enabling direct examination of resonantly enhanced nonlinear energy exchange in the following section.

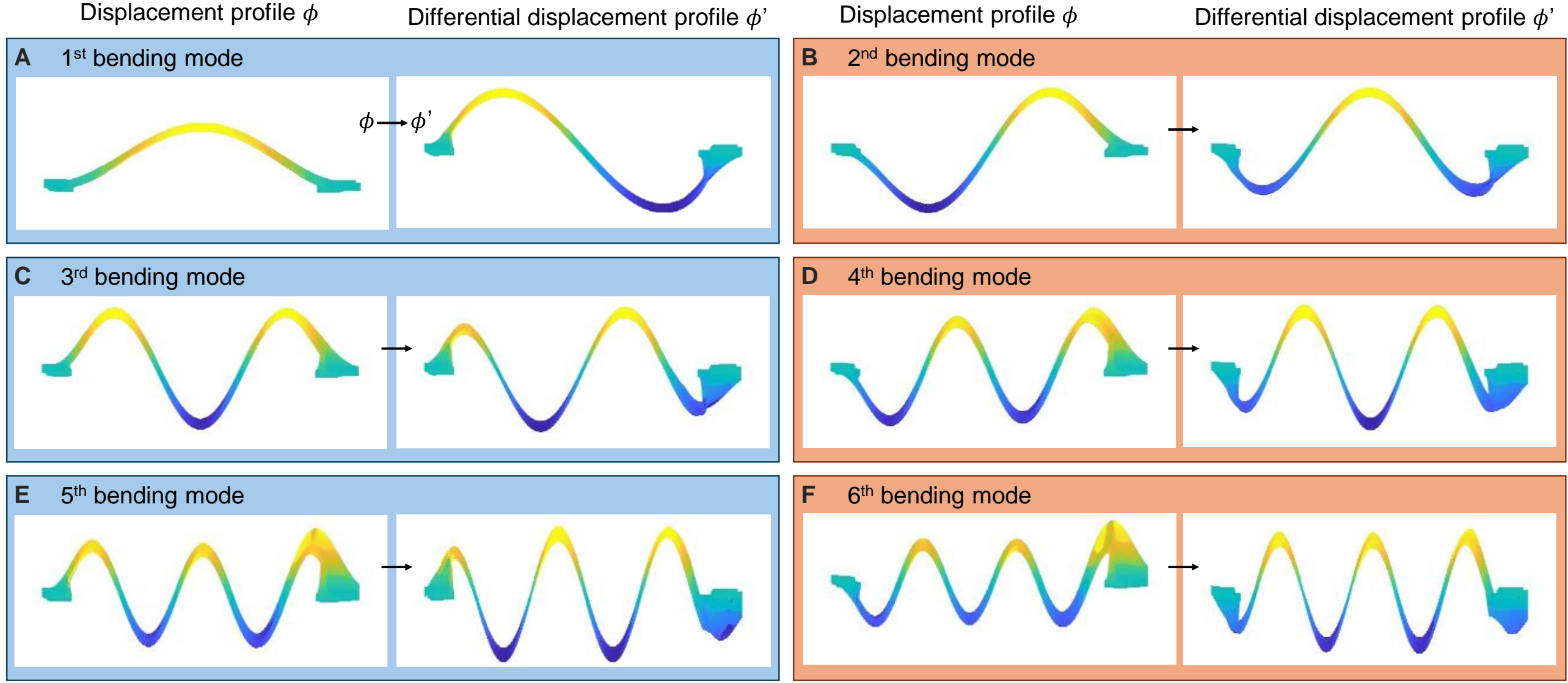

**Figure 2. Finite-element eigenmodes and corresponding slope distributions.** (**A**)-(**F**) show the displacement profile $\phi(u)$ and the corresponding spatial derivative $\phi'(u)$ of the first six bending modes. The modes separate into mirror-symmetric (1st, 3rd, 5th) and mirror-antisymmetric (2nd, 4th, 6th) classes with respect to the beam midpoint.

## Internal resonance visualized in real space

Figure 3 shows the nonlinear response of the fundamental bending mode in the vicinity of the commensurability condition $f_1 \approx f_3/5$. The illumination frequency in the phase-locked stroboscopic interferometry is set 1 Hz higher than the vibration frequency of the MEMS beam, so that the fast vibration motion is down-converted to a 1 Hz slow motion, captured by a CMOS camera [32]. We extract the spatially resolved amplitudes of the first and fifth harmonic components from the DIC contrast within the region marked by a red rectangle in Fig. 1A. The first- and fifth-harmonic responses are plotted as the black and red curves in Fig. 3A. Two reproducible suppression dips in the first-harmonic amplitude are observed at 358.9 kHz and 360.3 kHz. At both frequencies, the first-harmonic amplitude decreases while a fifth-harmonic component emerges, indicating energy transfer from the fundamental mode into a higher-order harmonic channel. Away from these dips, the motion is dominated by the first harmonic. The reproducibility of the dips across repeated frequency sweeps confirms that they correspond to stable internally resonant states rather than transient instabilities.

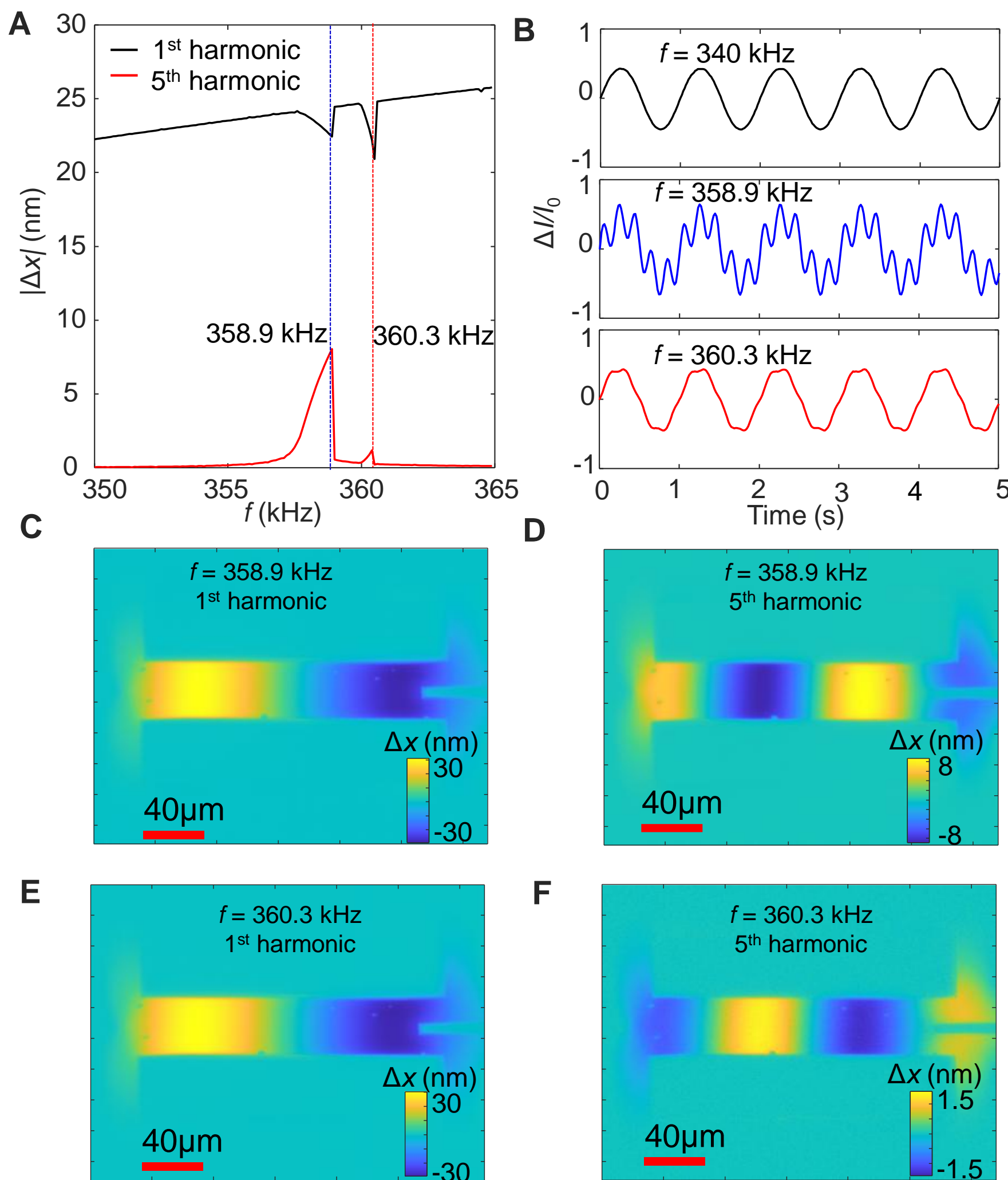

**Figure 3. Real-space visualization of 1:5 internal resonance.** (**A**) Nonlinear response of the fundamental bending mode in the vicinity of the commensurability condition $f_1 \approx f_3/5$, measured by phase-locked stroboscopic interferometry. The amplitudes of the first and fifth harmonic components, extracted from the DIC contrast within the region indicated in Fig. 1A, are plotted as the black and red curves, respectively. (**B**) Representative time-domain waveforms of the measured DIC contrast at an off-resonant frequency (340 kHz) and at the two suppression dips (358.9 kHz and 360.3 kHz), shown as the change in the light intensity $\Delta I$ normalized by the steady-state light intensity $I_0$. (**C**) The first-harmonic differential displacement map ($\Delta x$-map) at 358.9 kHz, whose spatial profile is characteristic of the fundamental bending mode. (**D**) The fifth-harmonic $\Delta x$-map, which exhibits three internal sign reversals along the beam length and matches the slope profile of the third bending mode. (**E,F**) First- and fifth-harmonic $\Delta x$-maps at 360.3 kHz, respectively.

To compare the dynamical states directly, Fig. 3B shows representative time-domain waveforms of the measured DIC contrast at an off-resonant frequency ($f$ = 340 kHz) and at two suppression dips ($f$ = 358.9 kHz, 360.3 kHz), shown as the change in the light intensity $\Delta I$ normalized by the steady-state light intensity $I_0$. As shown in Fig. 3B, off resonance, the waveform is nearly sinusoidal, consistent with motion confined to the fundamental mode. At both dips, the waveform becomes strongly distorted, reflecting superposition of first- and fifth-harmonic components. A key distinction between the two dips lies in the relative amplitude and phase of the fifth harmonic. At 358.9 kHz (left dip), the fifth-harmonic amplitude reaches approximately 30% of the first harmonic, producing pronounced waveform distortion. At 360.3 kHz (right dip), the fifth harmonic is substantially weaker (~5%), resulting in a milder distortion. Moreover, the harmonic components combine approximately in-phase at the left dip and out-of-phase at the right dip, revealing two distinct phase-locked steady states of the internal resonance. The real-space redistribution of vibrational motion during internal resonance is further illustrated in **Supplementary Movie 1.**

To resolve the spatial structure of the interacting modes, we reconstruct harmonic-resolved differential displacement maps ($\Delta x$-maps) at on-resonance frequencies. The stroboscopic DIC measurement provides the spatial difference of out-of-plane displacement between adjacent points separated by a fixed interval along the beam axis. In our experiment, the interval is ~1.3 μm, which is much smaller than the beam dimensions (length $L$ = 200 μm, width $W$ = 40 μm). Under this condition, the differential displacement is proportional to the spatial derivative $\phi'$ of the mode shape shown in Fig. 2, preserving nodal structure and mirror symmetry without requiring numerical integration.

Figure 3C shows the first-harmonic $\Delta x$-map at the in-phase internal resonance frequency (358.9 kHz). The spatial profile exhibits the single sign reversal characteristic of the fundamental bending mode. In contrast, the fifth-harmonic $\Delta x$-map measured at the same frequency (Fig. 3D) displays three internal sign reversals along the beam length, matching the slope profile of the third bending mode identified in Fig. 2C. The emergence of this multi-node structure directly visualizes the third bending mode carrying the generated fifth harmonic during internal resonance.

Figures 3E and 3F show the first- and fifth-harmonic $\Delta x$-maps at 360.3 kHz. The spatial patterns are nearly identical to those observed at 358.9 kHz, confirming that the same pair of eigenmodes participates in the interaction. The key distinction is a global phase inversion of the fifth-harmonic component relative to the first harmonic, corresponding to the out-of-phase steady state.

Together, these measurements provide direct real-space evidence that near-integer frequency matching activates coherent nonlinear energy exchange between distinct eigenmodes: suppression of the fundamental response coincides with the appearance of a fifth-harmonic component whose spatial profile corresponds to the third bending mode. The existence of two suppression dips with different fifth-harmonic amplitudes and opposite phase-locked character is consistent with multiple stable steady-state solutions of the 1:5 internally resonant dynamics.

## Off-resonant mixing reveals parity selection

When frequency commensurability is satisfied, the energy exchange is strongly enhanced through internal resonance. However, such observations alone cannot determine whether a generated harmonic is carried by another eigenmode only when resonantly amplified, or whether modal pathway selection persists away from frequency matching. To distinguish between these possibilities, we next examine off-resonant nonlinear mixing, where the role of spatial symmetry can be tested independently of integer frequency ratios.

To this end, the MEMS beam was driven outside the internal resonance regime at a drive frequency of $f = 340$ kHz. Under this condition, the system remains in the strongly nonlinear Duffing regime of the fundamental mode but avoids any integer frequency matching with higher-order modes. Using the same phase-locked multi-harmonic DIC analysis employed in the internal resonance measurements, we reconstructed differential displacement maps for individual harmonic components from the distributed DIC contrast. Figures 4A–B present the $\Delta x$-maps at the first and third harmonic frequencies, respectively. For clarity, line-cuts along the beam axis, indicated by the red arrows in Figs. 4A–B, are plotted in Figs. 4C–D.

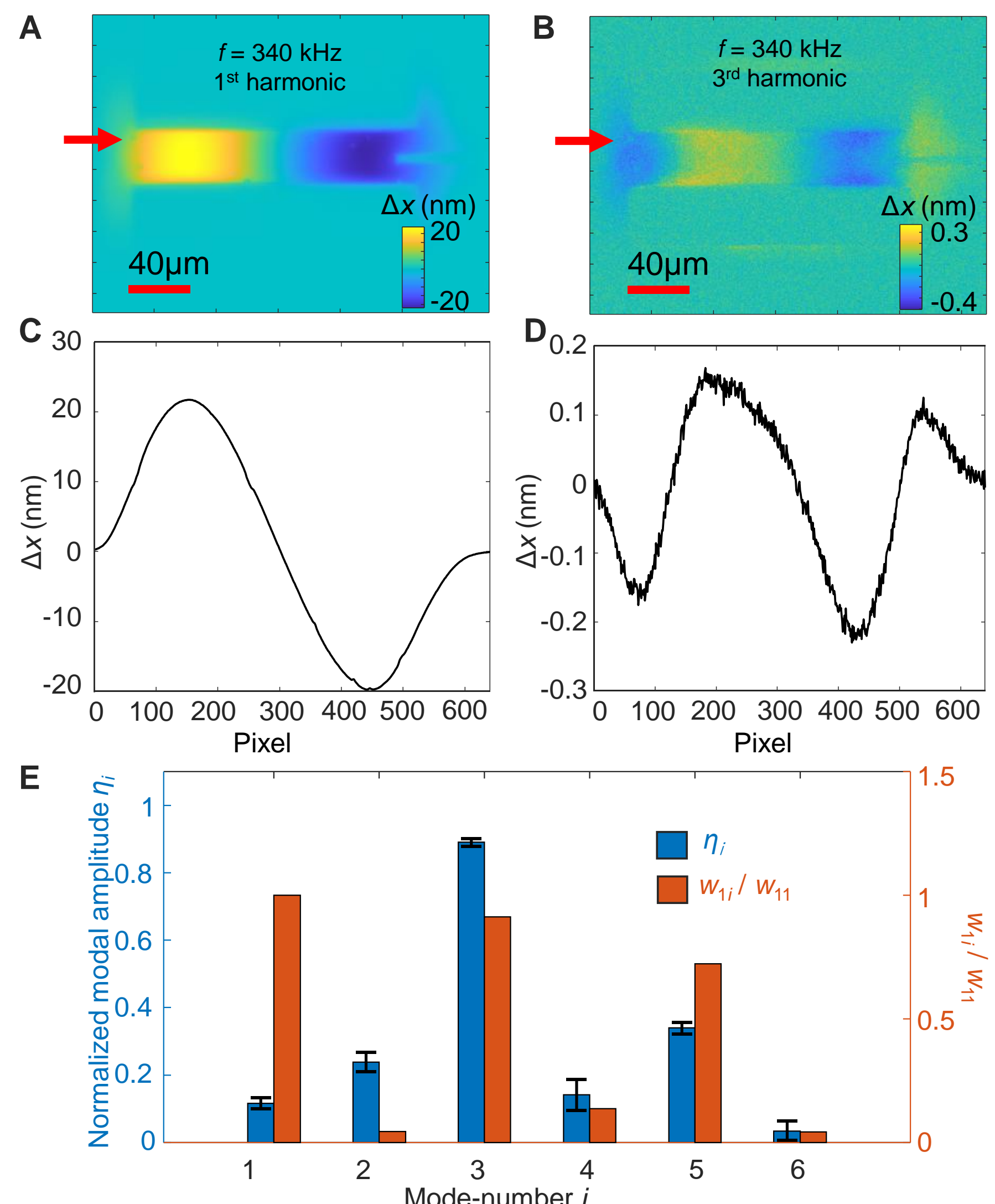

**Figure 4. Off-resonant nonlinear mixing reveals parity-selected nonlinear coupling.** (**A**,**B**) Harmonic-resolved differential displacement maps (Δ*x*-maps) measured at the first and third harmonic frequencies, respectively, with the MEMS beam driven at $f$ = 340 kHz outside the internal resonance regime. (**C**,**D**) Line cuts of the Δ*x*-maps in (**A**) and (**B**) taken along the beam axis, as indicated by the red arrows. (**E**) The blue and red bars show the normalized modal amplitude of each mode ($\eta_i$) in the third-harmonic motion and the calculated normalized modal overlap factors $w_{1i}/w_{11}$ ($i$ = 1–6), respectively. Error bars on the blue bars indicate 3σ-equivalent Monte Carlo uncertainties.

As shown in Figs. 4A–D, the first harmonic differential displacement distribution agrees well with the spatial slope profile of the fundamental bending mode, confirming that the beam oscillates predominantly in the first bending mode at the drive frequency. In contrast, the third harmonic exhibits a more complex spatial structure, containing contributions from multiple eigenmodes. A notable feature is that the central region of the beam shows nearly zero differential displacement at the third harmonic. Because the DIC signal is proportional to the spatial derivative of the displacement profile, such a midpoint zero is characteristic of mirror-symmetric displacement modes. This observation suggests that the dominant modal components contributing to the third harmonic belong to the same parity class as the driven fundamental mode, indicating that spatial mode structure constrains the nonlinear energy-transfer pathway.

To quantitatively determine the modal contributions within each harmonic component, the measured differential displacement fields were projected onto normalized differential mode profiles obtained from FEM simulations of the eigenmodes shown in Fig. 2. A least-squares fitting procedure was used to extract the differential displacement amplitude ($D_i$) of the $i^{\text{th}}$ bending mode from the third harmonic motion. The normalized modal amplitude $\eta_i$ for the third-harmonic motion is defined as

$$\eta_i = \frac{|D_i|}{\sqrt{\sum_j D_j^2}}. \tag{1}$$

Details of the modal decomposition are provided in **Supplementary Note 1**.

The extracted normalized modal amplitudes $\eta_i$ are summarized by the blue bars in Fig. 4E. The third harmonic contains significant contributions from the third and fifth bending modes, while the contributions from the second and fourth bending modes are small. This modal hierarchy is nontrivial. The third harmonic frequency is $3f$ = 1.02 MHz, which lies much closer to the second bending mode (892 kHz) than to the third bending mode (1.735 MHz). If frequency proximity alone determined which mode carries the generated harmonic, one would expect a stronger excitation of the second bending mode. However, the normalized amplitude of the second bending mode ($\eta_2 = 0.21$) is much smaller than that of the third bending mode ($\eta_3 = 0.91$). Similarly, the normalized amplitude of the fourth bending mode ($\eta_4 = 0.12$) is also much smaller than that of the fifth bending mode ($\eta_5 = 0.31$) even though its frequency is much closer to the third-harmonic frequency.

The results in Fig. 4 clearly indicate that, even away from integer frequency matching, the generated harmonic motion can be dominated by spatial eigenmodes that are not the nearest in frequency. Instead, modes sharing the driven mode's spatial symmetry are preferentially activated through nonlinear coupling. In the present beam, the first, third and fifth bending modes (odd modes) possess mirror-symmetric shapes, whereas the second, fourth and sixth modes (even modes) are mirror-antisymmetric. The observed dominance of odd–odd coupling and suppression of odd–even interaction provide direct experimental evidence that spatial parity, rather than frequency proximity alone, selects nonlinear modal pathways.

## Reciprocal excitation confirms parity selection

To further confirm this parity-selection rule, the second bending mode was driven into its nonlinear oscillation regime at $f = 910$ kHz. The differential displacement maps of its first and third harmonic components are shown in Figs. 5A and 5B. Corresponding line-cuts along the beam axis are presented in Figs. 5C and 5D, and the extracted normalized modal amplitudes $\eta_i$ are summarized by the blue bars in Fig. 5E. The blue bars in Fig. 5E show that, when the second mode (even mode) is nonlinearly excited, its third harmonic is predominantly composed of the fourth bending mode (even mode). In contrast, contributions from the first and third bending modes (odd modes) remain strongly suppressed. This reciprocal behavior further confirms that the observed modal hierarchy is not specific to fundamental-mode excitation, but reflects parity-selective nonlinear coupling: nonlinear modal energy exchange occurs preferentially between modes of identical spatial symmetry.

Taken together, the observations in Figs. 3–5 demonstrate that nonlinear modal coupling in the present mechanical beam is fundamentally constrained by mirror symmetry. Cubic nonlinear interaction selectively transfers energy between modes with matching parity, while cross-parity coupling is strongly suppressed. The results establish spatial parity as a selection rule underlying both internal resonance and off-resonant nonlinear modal mixing.

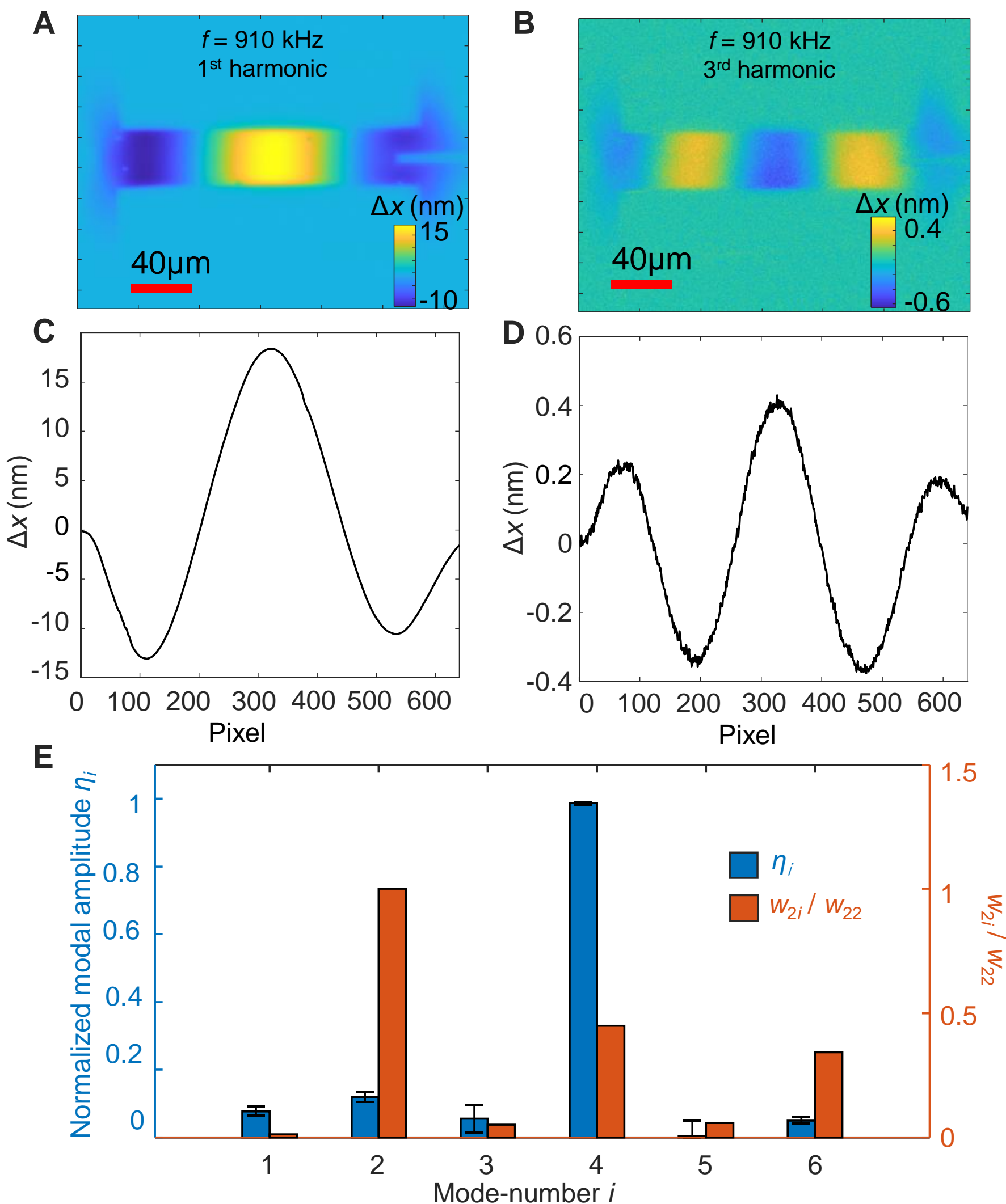

Figure 5. **Reciprocal verification of parity-selected nonlinear coupling using excitation of the second bending mode driven at *f* = 910 kHz in the nonlinear regime.** (**A,B**) Harmonic-resolved differential displacement maps ($\Delta x$-maps) of the first and third harmonic components measured with the second bending mode driven into the nonlinear regime. (**C,D**) Line cuts of the $\Delta x$-maps in (**A**) and (**B**) taken along the beam axis. (**E**) The blue and red bars show the normalized modal amplitudes ($\eta_i$) in the third-harmonic motion, and the calculated normalized modal overlap factors $w_{2i}/w_{22}$ ($i$ = 1–6), respectively. Error bars on the blue bars indicate 3σ-equivalent Monte Carlo uncertainties.

## Nonlinear modal theory and symmetry selection rule

We use a theoretical model to explain the observed parity-dependent nonlinear mode coupling. To capture the experimentally observed energy exchange, we restrict the dynamics to two interacting modes: a strongly driven mode $d$ with a displacement $x_d(t)$ and a higher-order mode $h$ with a displacement $x_h(t)$, such that

$$X(u,t) = x_d(t)\phi_d(u) + x_h(t)\phi_h(u) \tag{2}$$

where $\phi_d(u)$ and $\phi_h(u)$ are the mode-shape functions of the two modes.

A detailed derivation of the nonlinear modal equations, starting from the Euler–Bernoulli beam equation with vibration-induced dynamic tension, is provided in **Supplementary Note 2**. When the nonlinearity of beam motion is considered up to the third order, the equations of motion for the two modes take the generic form [4, 33, 34]:

$$\ddot{x_d} + \omega_d^2 x_d + \left(\frac{E}{2\rho L^2} w_{dd}^2\right) x_d^3 + \frac{E w_{hd} w_{hh}}{2\rho L^2} x_h^3 + \left(\frac{3E}{2\rho L^2} w_{dd} w_{hd}\right) x_h x_d^2 + \frac{E\left(2w_{hd}^2 + w_{dd} w_{hh}\right)}{2\rho L^2} x_d x_h^2 = F\cos(\Omega t), \tag{3}$$

$$\ddot{x_h} + \omega_h^2 x_h + \left(\frac{E}{2\rho L^2} w_{hh}^2\right) x_h^3 + \frac{E w_{hd} w_{dd}}{2\rho L^2} x_d^3 + \left(\frac{3E}{2\rho L^2} w_{hh} w_{hd}\right) x_d x_h^2 + \frac{E\left(2w_{hd}^2 + w_{dd} w_{hh}\right)}{2\rho L^2} x_h x_d^2 = 0. \tag{4}$$

where $\omega_d$ and $\omega_h$ are the resonance frequencies of the two modes; $\rho$ is the density; $E$ is the Young's modulus; $L$ is the beam length; $F$ is the driving amplitude, and $\Omega$ is the angular drive frequency; $w_{ij}$ denotes a modal-overlap parameter that determines the nonlinear coefficient and mode coupling coefficient, as

$$w_{ij} = \int_0^L \frac{\partial \phi_i}{\partial u}\frac{\partial \phi_j}{\partial u} du \quad (i,j = d,h) \tag{5}$$

Equations (3) and (4) show that mode $d$ is driven by the external force, whereas mode $h$ is excited through nonlinear coupling to the harmonic components generated by the driven-mode motion. Although the diagonal factors $w_{dd}$ and $w_{hh}$ set the self-nonlinearities of the individual modes, the resonant energy-transfer terms are governed by the cross-overlap factor $w_{hd}$ (= $w_{dh}$). This cross-overlap factor is symmetry dependent and therefore establishes a symmetry-selected energy-routing pathway.

The red bars in Fig. 4E and Fig. 5E show the normalized overlap factors $w_{di}/w_{dd}$ for the two driven-mode cases, $d = 1$ and $d = 2$, respectively, calculated from the numerical mode shapes shown in Fig. 2. The diagonal terms $w_{11}$ and $w_{22}$ are self-overlap factors of the driven modes; they quantify the self-Duffing nonlinearity and serve as normalization references, rather than representing intermodal transfer into modes 1 or 2. They should therefore not be interpreted as predicting large $\eta_1$ in Fig. 4E or large $\eta_2$ in Fig. 5E. Instead, the parity selection should be read from the off-diagonal overlaps. For first-mode excitation, the same-parity channels $w_{13}$ and $w_{15}$ are large, whereas opposite-parity channels such as $w_{12}$, $w_{14}$, and $w_{16}$ are strongly suppressed. For second-mode excitation, the same-parity channels $w_{24}$ and $w_{26}$ are large, whereas opposite-parity channels such as $w_{21}$, $w_{23}$, and $w_{25}$ remain small. This off-diagonal overlap hierarchy is consistent with the experimentally observed parity hierarchy in the third-harmonic modal amplitudes $\eta_i$.

For an ideally mirror-symmetric structure, the overlap integral $w_{ij}$ vanishes identically when the two modes have opposite parity. As a result, the coupling terms responsible for coherent energy exchange vanish, and the remaining cross-nonlinear terms, $x_d x_h^2$ in Eq. (3) and $x_h x_d^2$ in Eq. (4), act predominantly as dispersive terms [33, 34]. The primary effect of such dispersive terms is to produce amplitude-dependent stiffness corrections, leading to frequency renormalization rather than coherent energy transfer [33, 34]. The model therefore explains the experimentally observed hierarchy: an odd driven mode preferentially activates odd higher-order modes, whereas an even driven mode activates even higher-order modes. Opposite-parity channels remain suppressed even when they are spectrally closer to the generated harmonic, establishing spatial parity as a selection rule for nonlinear modal energy routing.

## Discussion

These findings show that a generated harmonic is not simply directed to the nearest spectral resonance; instead, spatial symmetry selects the modal-coupling channels through which nonlinear energy exchange occurs. Frequency commensurability determines when an existing energy-routing channel can be amplified through internal resonance, whereas spatial symmetry determines which channels are intrinsically allowed or suppressed. Thus, internal resonance is not simply a spectral matching condition; it is the resonant enhancement of a symmetry-selected energy-routing pathway.

Complete suppression of parity-forbidden coupling is expected only in ideally mirror-symmetric structures. In real devices, weak fabrication asymmetry makes opposite-parity overlap integrals small but finite. Such weak channels may still become observable when frequency commensurability resonantly amplifies them, explaining why nominally symmetry-forbidden internal resonances can appear in experiments [9, 17]. These channels, however, originate from incidental symmetry breaking and are therefore expected to be weak and sample-dependent, in contrast to robust symmetry-allowed coupling channels [10].

By combining real-space harmonic imaging, off-resonant control experiments and modal-overlap theory, our work identifies spatial symmetry as a design principle for selecting and controlling nonlinear energy flow. This framework can be extended beyond MEMS resonators to multimode nonlinear wave systems in which mode shape, symmetry and frequency matching can be engineered to select nonlinear energy-routing pathways, including NEMS devices, optomechanical platforms and phononic structures.

## Materials and Methods

### Device fabrication and mechanical characterization

The devices were fabricated from a (100)-oriented silicon-on-insulator (SOI) wafer consisting of a 2-μm device Si layer, a 3-μm buried $SiO_2$ layer, and a 450-μm handle Si layer, with a resistivity exceeding 2000 Ω·cm. The beam geometry was defined by anisotropic wet etching of the device layer using tetramethylammonium hydroxide (TMAH). The buried oxide layer was subsequently removed by selective etching in diluted hydrofluoric (HF) acid to release the doubly clamped Si beam.

For mechanical measurements, the chip was mounted on a piezoelectric actuator inside a vacuum chamber maintained at approximately 0.1 Pa to minimize air damping. An AC driving voltage $V_{AC}$ was applied to the piezoelectric actuator to excite out-of-plane beam vibration. The oscillatory motion was characterized either by an LDV combined with a lock-in amplifier for frequency-domain measurements or by a stroboscopic DIC microscope for spatially resolved imaging.

### Stroboscopic DIC imaging and harmonic decomposition

In stroboscopic DIC microscopy, the MEMS beam oscillating at a fundamental frequency $f$ is illuminated by a pulse-driven blue LED (wavelength $\lambda = 465$ nm, duty cycle ≈ 10%) at a slightly detuned frequency $f + 1$ Hz. Owing to the stroboscopic effect, the high-frequency mechanical oscillation is down-converted into an apparent slow modulation at 1 Hz, enabling phase-resolved imaging of fast vibration. The spatially resolved differential displacement $\Delta X(u,t)$ is extracted from changes in the DIC contrast $I(u,t)$ from

$$I(u,t) = I_0(u,t)\left(1 + \cos\left(\frac{4\pi}{\lambda}\Delta X(u,t) + \frac{\pi}{2}\right)\right), \quad \text{(M1)}$$

$$\Delta X(u,t) = \frac{\lambda}{4\pi} \times \arcsin\left(1 - \frac{I(u,t)}{I_0(u,t)}\right), \quad \text{(M2)}$$

where $I_0(u,t)$ denotes the steady-state light intensity observed without vibration.

When multiple harmonic components are present in the motion, the displacement field can be expressed as

$$\Delta X(u,t) = \sum \Delta x_i(u)\cos(2\pi i f_{\text{beat}} t + \theta_i) \quad \text{(M3)}$$

where $\Delta x_i(\text{u})$ denotes the spatially distributed amplitude, $\theta_i$ is a global phase shift of each harmonic that is independent of position, and $f_{\text{beat}}$ = 1 Hz is the down-converted frequency. To recover a signed spatial distribution, a global phase shift $\theta_0$ is applied to the reference signal $\cos(2\pi i f_{\text{beat}} t + \theta_0)$ during demodulation such that the extracted amplitude is maximized. This effectively removes the common phase $\theta_i$, yielding a real-valued signed amplitude map ($\Delta x$-map).

Because the measured quantity is the displacement difference between adjacent sheared points rather than the absolute out-of-plane displacement, the relevant optical phase excursion is substantially reduced, which significantly improved the allowed displacement range. The phase-based conversion further corrects the dominant nonlinear intensity response, giving linear readout of the vibration motion, and subsequently reliable modal decomposition.

## Modal decomposition

Modal decomposition was performed directly on the harmonic-resolved differential displacement maps obtained from the DIC measurements. Because the DIC signal corresponds to the displacement difference between two adjacent points along the beam axis rather than the absolute out-of-plane displacement, the measured maps were fitted using normalized differential mode profiles from FEM simulations. Error bars represent the 3σ-equivalent Monte Carlo uncertainty of the normalized fitted coefficients. The Monte Carlo analysis accounts primarily for uncertainty in the global horizontal alignment between the simulated differential mode profiles and the measured differential displacement maps; the fitting coefficients were recalculated for each randomly shifted realization. More details of the fitting procedure are provided in **Supplementary Note 1**.

## Data availability

The data that support the findings of this study are available from the corresponding author upon reasonable request. Source data for the main figures and Supplementary Movie will be made available in a public repository upon acceptance.

## Acknowledgements

We thank Dr. Hiromasa Shimizu for his support in the device fabrication process and fruitful discussions. The authors also acknowledge support from the NICT Advanced ICT Device R&D Promotion Center in device fabrication and characterization. This work was financially supported in part by the Adaptable and Seamless Technology transfer Program through Target-driven R&D (A-STEP) from Japan Science and Technology Agency (JST) Grant Number JPMJTR23R2, JSPS KAKENHI Grant Numbers 21K04151, 24K00937, and 26K22460.

## Author contributions

Ya Zhang conceived the idea of the work, performed data analysis, and wrote the manuscript. Yuko Terasawa performed the experiments and contributed to data analysis. Qian Liu fabricated the samples. Shumpei Takenaka contributed to the theoretical modeling. All authors participated in the discussions and reviewed the manuscript.

## Competing interests

The authors declare no competing interests.